\documentclass[aps,pre,preprint,showpacs,floatfix,unsortedaddress,
superscriptaddress]{revtex4-1}

\usepackage{epsfig}
\usepackage{float}
\usepackage{latexsym}
\usepackage{amsmath}
\usepackage{amsfonts}
\usepackage{amssymb}
\usepackage{amsbsy}
\usepackage{subfigure}
\usepackage{bm}
\usepackage{color}
\usepackage{dcolumn}
\usepackage{hyperref}

\graphicspath{{./figures/}}

\begin{document}
\title{Pattern Formation in Liquid-Vapor  
Systems under Periodic Potential and Shear}
\author{A. Coclite}
\email[]{a.coclite@poliba.it}
\affiliation{
Dipartimento di Meccanica, Matematica e Management, Politecnico di Bari,
Via Re David 200, 70126 Bari, Italy}
\author{G. Gonnella}
\email[]{gonnella@ba.infn.it}
\affiliation{Dipartimento di
Fisica, Universit\`{a} di Bari,
 {\it and} INFN, Sezione di Bari,
Via Amendola 173, 70126 Bari, Italy}
\author{A. Lamura}
\email[]{a.lamura@ba.iac.cnr.it}
\affiliation{
Istituto Applicazioni Calcolo, CNR,
Via Amendola 122/D, 70126 Bari, Italy}
\date{\today}
\begin{abstract}
In this paper the phase behavior and pattern formation  in a sheared
nonideal fluid under a periodic potential  is studied. An
isothermal two-dimensional formulation of a lattice Boltzmann scheme
for a liquid-vapor system with the  van der Waals equation of state
is presented and validated. Shear is applied by moving walls and the
periodic potential varies along the flow direction. A region of
the parameter space, where in absence of flow a striped phase with
oscillating density is stable, will be considered. At low shear rates
the periodic patterns are preserved and slightly distorted by the
flow. At high shear rates the striped phase looses its stability and
travelling waves on the interface
between the liquid and vapor
regions are observed. 
These waves
spread over the whole system with
wavelength only depending on the length of the system.
Velocity field patterns, characterized by a single vortex, 
will be also shown.
\end{abstract}
\maketitle

\section{Introduction}

The phase transition behavior of liquid-vapor systems can be very
rich in presence of confining geometries or external fields
\cite{Bind08}. Recently, the case of fluids inside a static
potential periodically oscillating in one direction has been
carefully examined. The main result is the presence of a modulated
third phase, called ``zebra'' phase, with density varying between the
values of the vapor and the liquid phases. This phase, first found in
colloid-polymer mixtures and referred as laser induced condensation
\cite{Gotze03}, has been theoretically analyzed in
Refs.~\cite{Vink11,Vink12}. The  periodic potential can be realized in
$d=2$ dimensions by stripe patterned surfaces \cite{Gau} and in
$d=3$ by laser or electric fields. The zebra  phase can coexist
with the liquid or the vapor phase. The coexistence lines terminate
 at two critical points on one side and join  at a triple point
where liquid-vapor coexistence reappears. It is worth to remind that
in colloid-polymer mixtures the phase separation between a
colloid-rich and a colloid-poor phase can be described in terms of a
liquid-vapor transition \cite{poon02}. Therefore the new phenomena
induced by  the presence of oscillating external potential can be
relevant for a large class of systems.

In this paper we will consider the effects of an imposed flow on the
ordered patterns appearing in the zebra phase of a liquid-vapor
system. More specifically,  we will study a fluid confined between two
shearing walls with a
periodic potential varying  along the flow
direction. As it will be shown, at low shear rate the periodic
patterns are slightly distorted by the flow, as expected. New
interesting features appear at high shear rates when travelling waves on the
liquid-vapor interface appear.

In order to simulate the fluid system we use a lattice Boltzmann
model with a van der Waals equation of state.
Several approaches have been previously adopted to model
nonideal fluids by using the lattice Boltzmann equation (LBE). They
can be mainly divided in four groups: Color-gradient models
\cite{guns91,grun93}, pseudopotential models
\cite{shan93,shan94,shan06,yuan06,sbra07,sbra09bis,falc07,shan08,
zhan08,hyva08,kupe09,falc09,
sbra09,wang09,yu10,falc10,sbra11,colo12,li12}, free-energy models
\cite{swif95, swif96,inam00,bria04,wagn06,wagn07,li12bis},
and kinetic models 
\cite{he98,luo98,he99,luo00,he02,sofo04,noitermico,cris10,aiguo,phil12}.
The main difference among the cited models is related to the
procedure to introduce intermolecular force into multi-phase LBE.
In the color-gradient models the force is heuristically derived and
incorporated in the LBE via a forcing term; in the pseudopotential
models the force is derived from a potential and
enters the model through a modified velocity in the
equilibrium distribution functions (EDF) of the LBE. In the
free-energy models the EDF are redefined to include intermolecular
forces into the pressure tensor that fixes the second moment of the EDF, 
while in kinetic models the interparticle interaction is 
treated using a mean-field
approximation and is added to the LBE by a forcing term.
Recently, models have been proposed
where LBE contains a forcing term and EDF are redefined as well
\cite{he98,lee06,lee08,lee09,lee10,chia10,gros11}. Despite the
different ways to introduce the interaction force into the LBE, it
was shown that all the models can be recast in the form of a
standard LBE with a forcing term \cite{he98,luo00}. Models have
different pros and cons. For recent discussions the interested
reader may refer, among others, to
Refs.~\cite{falc11,guo11,lou12,li12}.

A big improvement in lattice Boltzmann modeling is the derivation of the
LBE on a more rigorous mathematical basis by a
Gauss-Hermite projection of the corresponding continuum equation
\cite{shan06bis}.
In this paper we aim at introducing a LBE model for liquid-vapor
systems consistent with such a systematic discretization of the Boltzmann
kinetic equation. 
Thermodynamics enters the model via a free-energy dependent term,
added as a body force to the LBE \cite{li07}, and a redefined
EDF as in Refs.~\cite{he98,lee06,lee08,lee09,lee10,chia10,gros11} so that
the fluid locally satisfies the van der Waals equation of state \cite{rowl82}.
This allows to avoid uncontrollable spurious terms in the continuum
equations.

The paper is organized as follows. In Sec.~II the LBE model is introduced
while validation tests are reported in Sec.~III. In Sec.~IV the effects
of shear on the morphology of the zebra phase are shown.

\section{Model} \label{sec:model}

We will consider in the following an isothermal model at temperature $T$
in $d=2$.
The evolution of the fluid is defined in terms of a set of $N$ discrete
distribution functions $\{f_i\}$ ($i=0,\ldots,N-1$) which obey the
dimensionless Boltzmann equation
\begin{equation}\label{evoleq}
f_{i}({\bf r}+{\bf e}_{i}\Delta t, t+\Delta t)-f_{i}({\bf r}, t)=
-\frac{\Delta t}{\tau}[f_{i}({\bf r}, t)-f_{i}^{eq}({\bf r}, t)]
+\Delta t {\cal F}_i ,
\end{equation}
where ${\bf r}$ and $t$ are the spatial coordinates and time, respectively,
$\{{\bf e}_{i}\}$ ($i=0,\ldots,N-1$) is the set of discrete
velocities, $\Delta t$ is the time step, $\tau$ is a relaxation
time, and ${\cal F}_i$ is a convenient forcing term to be
determined. The moments of the distribution functions
define the fluid density $n=\sum_i f_i$ and velocity ${\bf u}=\sum_i
f_i {\bf e}_i/n$.
The local EDF $\{f_i^{eq}\}$ ($i=0,\ldots,N-1$) are expressed by a
Maxwell-Boltzmann (MB) distribution.

Here we adopt a discretization 
in velocity space
of the MB distribution based on the
quadrature of a Hermite polynomial expansion of this distribution
\cite{shan06bis}. In this way it is possible to get a LBE that
allows to exactly recover a finite number of leading order
moments of the MB distribution.
In such a scheme it has been later realized
that a regularization step \cite{latt06} is needed for $\tau \neq 1$
to re-project the post-collision distribution functions onto the Hermite
space \cite{zhang06}. 
However, for practical purposes, such step is only  necessary  
when the Knudsen
number $Kn=\eta/(n \sqrt{T} L_y)$, where $\eta$ is the viscosity and $L_y$ the
system width, becomes larger than some value ($\approx 0.05$)
\cite{zhang06,colos10}. 
In the present
paper the value of $Kn$ is always smaller than 0.005 
so that the regularization step was not implemented.
The form of $f_i^{eq}$ will be such that
\begin{eqnarray}
\sum_i f_i^{eq} &=& n , \\
\sum_i f_i^{eq} {\bf e}_i &=& n {\bf u} ,\\
\sum_i f_i^{eq} {\bf e}_i {\bf e}_i &=& n T {\bf I}
+ n {\bf u} {\bf u} . \label{2nd}
\end{eqnarray}
The
velocity space is spanned by the leading Hermite polynomials
where the values of ${\bf e}_{i}$ are given by the abscissas of a
Gauss-Hermite quadrature in such a velocity space \cite{shan06bis}.
The local EDF 
at the second-order
Hermite expansion of the Maxwell-Boltzmann distribution are given by
\cite{shan06bis}
\begin{equation}\label{eqfunc}
f_i^{eq}({\bf r},t)=\omega_in\left[1+{\bf e}_i \cdot{\bf u}+
\frac{1}{2}{\bf uu:}({\bf e}_i{\bf e}_i-{\bf I})
+\frac{1}{2}(T-1)({\bf e}_i \cdot {\bf e}_i-2)\right]
\end{equation}
where ${\bf I}$ is the unitary matrix . The term proportional to
$(T-1)$ would normally vanish for isothermal systems with reference
temperature $T=1$ but in our case we will consider nonideal fluids
at different temperatures lower than the critical value $T_c$
which is chosen as a reference and set equal to 1 (see in
the following). On the two-dimensional square lattice with $N=9$ speeds
($D2Q9$) the previous second order expansion fixes the values $|{\bf
e}_i|=\sqrt{3}$ for horizontal and vertical links with $i=1 (East
(E) direction), 2 (North (N) )$, $3 (West (W)), 4 (South (S))$,
$|{\bf e}_i|=\sqrt{6}$ for diagonal links with $i=5 (NE)$, $6 (NW)$,
$7 (SW)$, $8 (SE)$, $|{\bf e}_0|=0$ for the rest velocity, and the
weights $\omega_i=1/9$ for $i=1-4$, $\omega_i=1/36$ for $i=5-8$,
and $\omega_0=4/9$.

In order to simulate the correct transport equations of a nonideal
fluid, we follow the approach proposed in Ref.~\cite{guo02} to
include the forcing term. This scheme was previously
applied to binary mixtures \cite{tiri09}, lamellar fluids
\cite{tiri11}, and nonideal fluids to investigate force imbalance \cite{guo11}
and force discretization \cite{lou12}.
Within this procedure, the velocity ${\bf u}$ in
(\ref{eqfunc}) is formally replaced by the physical velocity ${\bf
u}^*$ given by
\begin{equation}
n{\bf u}^*=\sum_i f_i {\bf e}_i +\frac{1}{2}{\bf F}\Delta t ,
\label{newvel}
\end{equation}
where ${\bf F}$ is the total force density acting on the fluid.
 ${\bf F}$ is the sum of the external forces  ${\bf F}_{ext}$
and of the intermolecular forces  ${\bf
F}_{int}$. In the next we will use
\begin{equation}
{\bf F}_{ext}=(A_0 \sin{(\frac{2 \pi x}{\lambda})}, 0)
\label{periodicforce}
\end{equation}
representing a static periodic force along the $x$-direction
of
amplitude $A_0$ and wavelength $\lambda$.
As we will show, ${\bf F}_{int}$ will be used to model a van der Waals fluid
described by the Navier-Stokes equation.

The forcing term ${\cal F}_i$ is a function of ${\bf F}$ and is
expressed as a power series at the second order
in the lattice velocities
\begin{equation}\label{forceterm}
{\cal F}_i=\omega_i\left[A+{\bf B}\cdot{\bf e}_i+
\frac{1}{2}{\bf C:}({\bf e}_i{\bf e}_i-{\bf I})\right] ,
\end{equation}
where $A$, ${\bf B}$, and ${\bf C}$ are functions of ${\bf F}$.

By using a second-order Chapman-Enskog expansion, we obtain the
continuity equation
\begin{equation}
\partial_tn+\partial_{\alpha}(n u_{\alpha}^*)=0
\label{conteqn}
\end{equation}
and the Navier-Stokes equation
\begin{equation}
\partial_t(n u_{\alpha}^*)+\partial_{\beta}(nu_{\alpha}^*u_{\beta}^*)=
-\partial_{\alpha} p^i + F_{int, \alpha} + F_{ext, \alpha}
+\partial_{\beta}\Big [ \eta(\partial_{\alpha}u_{\beta}^*
+\partial_{\beta}u_{\alpha}^*)\Big ] + o(u^{*3})
\label{nav}
\end{equation}
where $p^i=nT$ is the ideal pressure
and $\eta=n (\tau-\Delta t/2)$ the shear viscosity
(the bulk viscosity equals the shear viscosity in the present model),
provided the
following expressions for the terms $A$, ${\bf B}$, ${\bf C}$:
\begin{eqnarray}
A&=&0 , \\
B_{\alpha}&=&\left(1-\frac{\Delta t}{2\tau}\right)F_{\alpha} , \\
C_{\alpha \beta}&=&\left(1-\frac{\Delta t}{2\tau}\right)
\Big \{u^*_{\alpha} F_{\beta}+F_{\alpha} u^*_{\beta}+(1-T)\big [
u_{\alpha}^* \partial_{\beta} n
+u_{\beta}^* \partial_{\alpha} n
+ \partial_{\gamma} (n u_{\gamma}^*)\delta_{\alpha \beta} \big ] \Big \} 
\label{coeff}
\end{eqnarray}
are used.

We make a comment  here  about the Navier-Stokes equation (\ref{nav}).
Within the present formulation the only 
spurious term  appearing  in this equation is of order $u^{*3}$ 
and it will be  neglected. 
This is justified by the fact that it is $u^* \lesssim 0.1 $ 
(low Mach number limit) in all our simulations.
We recall that a second order
expansion of the EDF is used in this paper. Otherwise, in order to eliminate  
the spurious term of order $u^{*3}$,
we should have used
a third-order model $D2Q17$ \cite{shan06} requiring a larger number
of lattice velocities than the present model.
Other spurious  terms, that would have appeared in our model, 
proportional to derivatives of the local fluid velocity 
$u^*$ and of the density $n$, are 
exactly canceled by the proposed form of the quantity 
$C_{\alpha \beta}$ and  correspond to
the last bits multiplied
by $(1-T)$ in (\ref{coeff}).

In order to get the Navier-Stokes equation of a nonideal
fluid it has to be
\begin{equation}
F_{int, \alpha}=\partial_{\alpha} (p^i) - \partial_{\beta} \Pi_{\alpha \beta} .
\label{force}
\end{equation}
For a van der Waals fluid the pressure tensor ${\bf \Pi}$ can be derived 
from the free-energy
functional \cite{rowl82}
\begin{equation}
\Psi=\int d{\bf r} \Big [ \psi(n,T) +
\frac{\kappa}{2} (\nabla n)^2    \Big ]
\end{equation}
where the bulk free-energy density is
\begin{equation}
\psi=nT\ln \Big (\frac{3 n }{3 -n}\Big )-\frac{9}{8}n^2
\end{equation}
and the term proportional to  $\kappa$ expresses the energy cost for
the formation of interfaces and controls the surface tension. The
pressure tensor is then \cite {evan79}
\begin{equation}
\Pi_{\alpha \beta}=\Big [p^w - \kappa n \nabla^2 n -\frac{\kappa}{2} (\nabla n)^2
\Big ]
\delta_{\alpha \beta}+\kappa \partial_{\alpha} n \partial_{\beta} n
\label{prtensor}
\end{equation}
where
\begin{equation}
p^w=n\frac{\partial \psi}{\partial n}-\psi=\frac{3nT}{3-n}-\frac{9}{8}n^2
\label{eqstate}
\end{equation}
is the van der Waals equation of state with the critical point at
$n_c=1$ and $T_c=1$. 
The term $(3-n)/3$ at the r. h. s. of Eq.~(\ref{eqstate})
takes into account the excluded volume interaction while
the last term represents the attractive force between molecules.

The intermolecular force density then reduces to
\begin{equation}
F_{int, \alpha} = \partial_{\alpha} (p^i-p^w) + \kappa n \partial_{\alpha}(\nabla^2 n) .
\label{force2}
\end{equation}

The second moment of EDF (\ref{2nd}) is not changed to include the
effects of the pressure tensor as in the free-energy based model of
Ref.~\cite{swif96} 
since an internal force is adopted for this purpose.
We note that the presence of the forcing term in the LBE
does not allow to guarantee
local momentum conservation, differently from free-energy based models
where this leads to small spurious
velocities \cite{nour02}.  
However, our model, as shown, is free from unwanted
spurious terms in the continuum equations and has the relevant property 
of a phase diagram not depending
on the relaxation time
(see the discussion in the following Section).
Moreover, the surface tension is properly taken into account
into the pressure tensor (\ref{prtensor})
avoiding any lack of thermodynamic consistency.

When a shear flow is considered, moving walls have to be modeled in
order to enforce the flow. Here we place planar walls at the lower
and upper rows of the lattice setting a neutral wetting condition for
the density $n$ at the walls. This is obtained by imposing that
${\bf a}\cdot {\bf {\nabla}} n|_{walls}=0$, being ${\bf a}$ an
inward unit vector normal to the boundaries, which guarantees that
the angle between the walls and liquid-vapor interfaces 
is kept constant at the value $\pi/2$ radians. By using the approaches developed
in Refs.~\cite{lamu01,tiri11bis}, the expressions of the unknown
distribution functions at walls can be obtained. We discuss the case
of the lower wall (with similar considerations for the upper one).
After the propagation, the distribution functions $f_2(t)$,
$f_5(t)$, $f_6(t)$ are unknown. In order to have the fluid moving
with the wall velocity $-\dot{\gamma} L_y/2$ ($\dot{\gamma} L_y/2$ on
the upper wall), where $\dot{\gamma}$ is the shear rate and $L_y$ the
width of the system, it has to be
\begin{eqnarray}
f_0({\bf r},t)&=&\hat{n}-[f_1({\bf r},t)+f_3({\bf r},t)]-
2[f_4({\bf r},t)+f_7({\bf r},t)+f_8({\bf r},t)]
+\frac{\Delta t}{2}F_y\nonumber\\
f_5({\bf r},t)&=&f_7({\bf r},t)-\frac{1}{2}[f_1({\bf r},t)-f_3({\bf r},t)]-
\frac{\Delta t}{4}[F_x+F_y]-\frac{1}{4} n \dot{\gamma} L_y \nonumber\\
f_6({\bf r},t)&=&f_8({\bf r},t)+\frac{1}{2}[f_1({\bf r},t)-f_3({\bf r},t)]+
\frac{\Delta t}{4}[F_x-F_y]+\frac{1}{4} n \dot{\gamma} L_y \nonumber\\
f_2({\bf r},t)&=&f_4({\bf r},t)
\end{eqnarray}
where $f_0(t)$ is introduced as an independent variable \cite{lamu01} to impose
mass conservation and
\begin{equation}
\hat{n}=f_0(t-\Delta t)+f_4(t-\Delta t)+f_7(t-\Delta t)+
f_8(t-\Delta t)+f_1(t)+f_3(t)+f_4(t)+f_7(t)+f_8(t),
\end{equation}
with the quantities at time $(t-\Delta t)$ calculated at the previous
time step and not yet propagated over the lattice. 
Finally, the collision step is performed
over all the lattice sites, including the ones on the walls.

The spatial derivatives in (\ref{force2})
are calculated using
a general second-order finite difference scheme in order
to ensure a higher isotropy \cite{pool08}, which helps in reducing
spurious velocities at interfaces \cite{shan06,sbra07}.
The schemes for the $\partial_x$ and the $\nabla^2$ operators
are, respectively,
\begin{equation}
\partial_x  =
\frac{1}{\Delta x}
\left[
\begin{array}{ccc}
-M & 0 & M \\
-N & 0 & N \\
-M & 0 & M \\
\end{array}
\right]
\label{xstencil}
\end{equation}
\begin{equation}
\nabla^2   =
\frac{1}{{\Delta x}^2}
\left[
\begin{array}{ccc}
R & Q & R \\
Q & -4\left(Q+R\right) & Q \\
R & Q & R \\
\end{array}
\right]
\label{laplstencil}
\end{equation}
where the relationships $2N+4M=1$ and $Q+2R=1$ must hold for the
discrete derivatives to be consistent with the continuous ones
\cite{pool08}. $\Delta x$ is the lattice space unit. In
(\ref{xstencil})-(\ref{laplstencil}) the central entry corresponds
to the lattice point where the derivative is meant to be computed,
and the other entries refer to the eight neighbor lattice sites. The
derivatives of the density $n$ are computed by summing up
all the values of $n$ in the nine entries on the lattice with the
weights in the matrices (\ref{xstencil})-(\ref{laplstencil}). The
$y$ derivative is computed by transposing the matrix
(\ref{xstencil}). The free parameters $N$ and $Q$ will be chosen to
minimize either the spurious velocities at interfaces (optimal
velocity choice - OVC) or the deviation of the equilibrium fluid
densities from the theoretical values in the phase diagram (optimal
density choice - ODC). The values $N=1/2$, $M=0$, $Q=1$, and $R=0$
correspond to the standard central difference scheme denoted as standard
choice (SC).
We will compare SC, OVC, and ODC in the next Section.

\section{Validation}

\begin{figure}[H]
\begin{center}
\includegraphics*[width=.45\textwidth,angle=270]{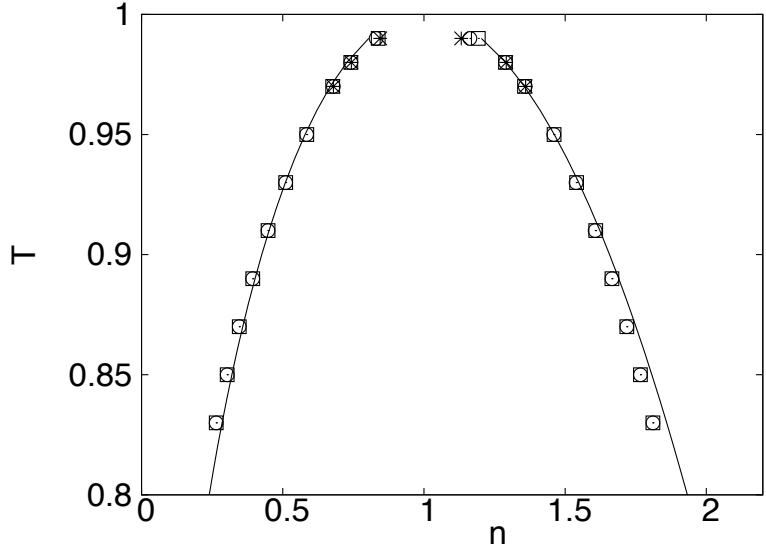}
\caption{The equilibrium phase diagram in the plane $(n,T)$ 
is shown for the SC case
($\ast$), the ODC case ($\Box$), and the OVC case ($\circ$) with
$\kappa=0.3$ and $\tau=1$. Temperature and density are
expressed in units of the critical values. The continuous
line represents the theoretical prediction obtained from the Maxwell
construction. Temperature and density are measured in units
of $T_c$ and $n_c$, respectively.} \label{fig:phase_diag}
\end{center}
\end{figure}

In all the simulations the values $\Delta x=1$ and
$\Delta t=\sqrt{3}/3$ were used to fix $|{\bf e}_i|=\Delta x/\Delta
t=\sqrt{3}$ for $i=1-4$, as prescribed by the Gauss-Hermite
quadrature on the $D2Q9$ lattice. 
In the following all the quantities will be expressed, according
to their dimensions, in units
of $\Delta x$, $\Delta t$, $n_c$, and $T_c$.
A liquid droplet of density
$n_L^{th}$ with radius $15 \Delta x$ was initialized with a sharp
interface in a vapor environment of density $n_V^{th}$ with periodic
boundary conditions along both the spatial directions with ${\bf
F}_{ext}=0$. The values $n_L^{th}$ and $n_V^{th}$ depend on the
chosen temperature $T$ and were taken from the theoretical phase
diagram obtained by using the Maxwell construction. The numerical
phase diagram was obtained by letting the liquid droplet to relax
and observing the equilibrium values of the density inside and outside
the droplet. The parameter $\kappa$ was taken to be $0.3$ in order
to have an equilibrium interface of about $6$ lattice spacings (see
the following) and the relaxation time was $\tau=1$. The obtained
phase diagram is shown in Fig.~\ref{fig:phase_diag}. In the SC case
the droplet keeps its circular shape down to $T \simeq 0.97$, while
for lower values of $T$ it is deformed into a square-like droplet. 
A wider temperature range
where the model can operate, also preserving the droplet circular
shape, is observed when using either the ODC or the OVC scheme for
fixing the values of $N$ and $Q$ in
(\ref{xstencil})-(\ref{laplstencil}). In these latter cases the
system was simulated by varying $N$ in the range $[0,2]$ and $Q$ in
$[0,4]$.  In the ODC the optimal couple of values $(N,Q)_{ODC}$ was
chosen in such a way to minimize the quantity
$(|n_L^{th}-n_L|+|n_V^{th}-n_V|)$ where $n_L$ and $n_V$ are the
numerical liquid and vapor densities at equilibrium, respectively.
In the OVC the optimal couple $(N,Q)_{OVC}$ was such to minimize the
maximum value of the spurious velocities observed at the droplet
interface at equilibrium. The presence of unphysical velocities at
interfaces is a notorious problem in LBE models of
multi-phase/component fluids. In the case of nonideal systems
several procedures have been devised in order to face such a problem
\cite{cris03,wagn03,shan06,sbra07,wagn07,pool08}. Here we adopt the
idea of increasing the isotropy of spatial numerical derivatives
\cite{shan06,sbra07,pool08}.
The optimal values of the ODC and OVC cases are reported in Table
\ref{table}. From the numerical results it appears that the OVC
scheme gives an error in the equilibrium densities comparable with
the one obtained by the ODC scheme 
(see Fig.~\ref{fig:phase_diag}). The present model
is capable to handle with enough accuracy density ratios 
$n_L/n_V \simeq 7$.

A comment is here in order about the numerical stability of the
model which limits the maximum value $n_L/n_V$ that can  be
simulated. We aimed to model a
nonideal fluid described by the van der Waals equation of state
which takes into account the volume excluded by fluid particles. 
As a consequence, the speed of sound
$c_s^2=\frac{\partial p^w}{\partial n}=\frac{9T}{(3-n)^2}
-\frac{9}{4}n$ increases when
both decreasing $T$ and approaching the liquid branch of the phase
diagram. When $c_s$ is larger than the lattice speed $\sqrt{3}$ (at
$T=0.8$ it is $c_s \simeq 1.97$ on the liquid branch), the numerical
stability is compromised and, for example, some assumptions on the
nonideal equation of state have to be used to circumvent the problem
\cite{colo12}. Here we tried to follow the approach proposed in
Ref.~\cite{wagn07} by multiplying the pressure tensor ${\bf \Pi}$ by
a numerical factor $\zeta$ which does not alter the phase diagram of
the model, hoping in an improvement of  stability. We did
a linear stability analysis in a one-dimensional system
\cite{kupe10},
finding the values of $\zeta$, as a function
of $\kappa$ and $T$, satisfying the stability condition. For these
values of $\zeta$ the model has a stability range in $T$ wider than
in the SC case but still narrower with respect to the ODC and OVC
cases thus making this approach not really useful. It might  well be
that the one-dimensional 
linear stability analysis is not enough to give the proper
values of $\zeta$ to increase the numerical stability in $d=2$.
Further work will be
necessary to better understand this issue.

\begin{figure}[H]
\begin{center}
\includegraphics*[width=.45\textwidth,angle=270]{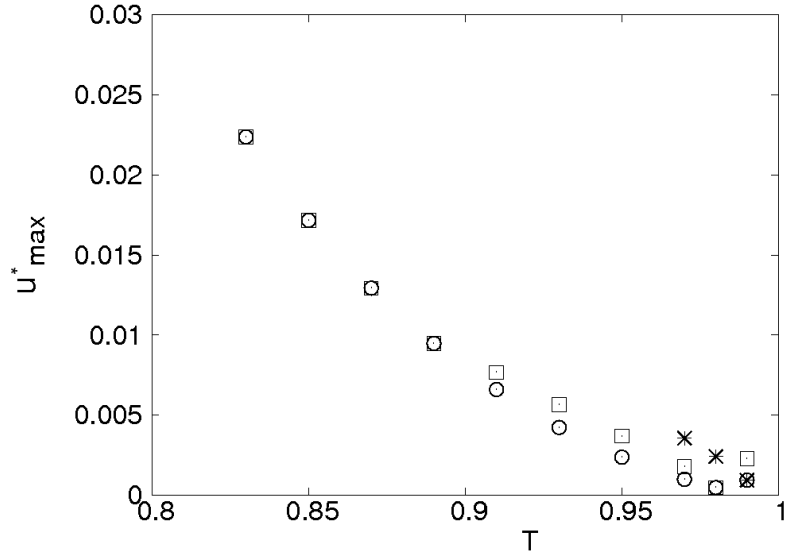}
\caption{The maximum value of the spurious velocities for an equilibrated
droplet is shown as a function of temperature $T$
for the SC case ($\ast$),
the ODC case ($\Box$), and the OVC case ($\circ$) with $\kappa=0.3$ and
$\tau=1$. Velocity is measured in units of $\Delta x / \Delta t$ and
temperature in units of $T_c$.
\label{fig:spvel_T}}
\end{center}
\end{figure}

The values of the maximum spurious velocities $u^*_{max}$ for the SC, ODC, and
OVC cases are shown in Fig.~\ref{fig:spvel_T} as a function of temperature.
It appears that $u^*_{max}$ is minimized in the OVC case, as expected, and
decreases when increasing the temperature. For this reason the OVC was
preferred and adopted in the following.

\begin{figure}[H]
\begin{center}
\includegraphics*[width=.49\textwidth]{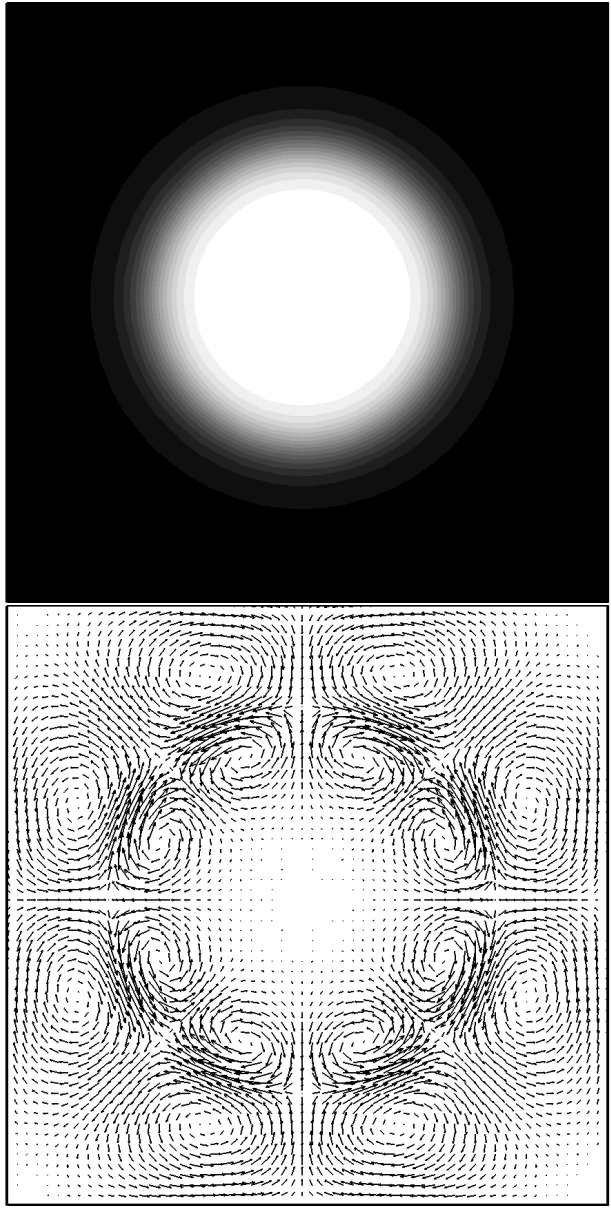}
\caption{(Upper panel) The density contour of an equilibrated
droplet with radius of 15 lattice units
is shown for $\kappa=0.3$ and $\tau=1$  at $T=0.95$ in the OVC case.
Grey-scaling from black $\rightarrow$ white corresponds
to vapor $\rightarrow$ liquid.
(Lower panel) The corresponding velocity pattern.
\label{fig:contour}}
\end{center}
\end{figure}

\begin{figure}[!h]
\begin{center}
\includegraphics*[width=.45\textwidth,angle=270]{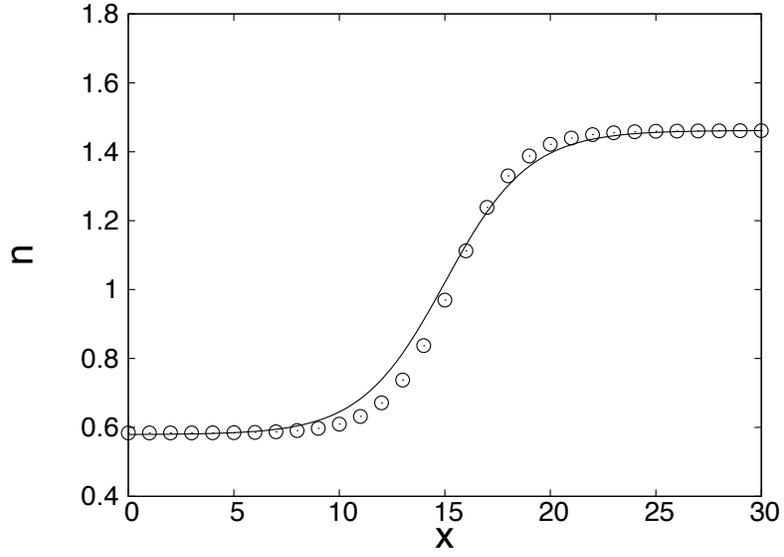}
\caption{The density profile across the interface of the
droplet of Fig.~\ref{fig:contour}.
The line
represents the prediction (\ref{interface}).
The measured values for density in the bulk
are $n_L=1.462721$ and  $n_V=0.580015$ to be compared with the
theoretical values  $n_L^{th}=1.461727$ and  $n_V^{th}=0.579015$.
Density is measured in units of $n_c$ and space in units of $\Delta x$.
\label{fig:interf}}
\end{center}
\end{figure}

The equilibrium density pattern of the liquid droplet at $T=0.95$
with the corresponding velocity field is reported for the OVC case
in Fig.~\ref{fig:contour}. Velocities are not zero
along the interface. The density profile across a section of the
droplet is represented in Fig.~\ref{fig:interf} with an interface of
about $6$ lattice spacings smoothly
interpolating between the liquid and vapor states. The profile is
well described by the approximate solution \cite{wagn07}
\begin{equation}
n(x)=n_V^{th}+\frac{(n_L^{th}-n_V^{th})}{2}\Big[1+\tanh \Big (\frac{x}{\sqrt{2\kappa/(1/T-1)}}
\Big ) \Big ]
\label{interface}
\end{equation}
of the equation $\partial_{\alpha} \Pi_{\alpha \beta}=0$.

\begin{figure}[H]
\begin{center}
\includegraphics*[width=.45\textwidth,angle=270]{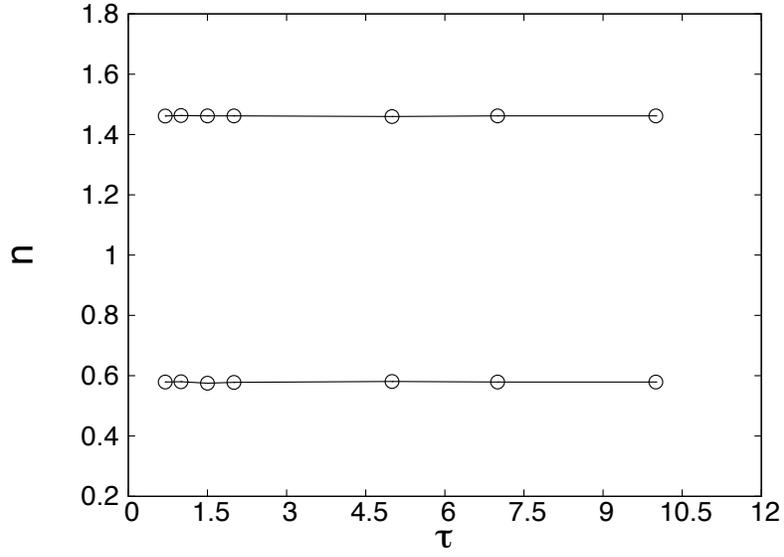}
\caption{The equilibrium densities are shown as a function of the
relaxation time $\tau$ for $\kappa=0.3$ at $T=0.95$ in the OVC case.
Continuous lines are a guide to the eye.
Density variations are less than $1\%$. Explicit values are given
in Fig.~\ref{fig:interf}.
Density is measured in units of $n_c$ and time in units of $\Delta t$.
\label{fig:dens_tau}}
\end{center}
\end{figure}

\begin{figure}[H]
\begin{center}
\includegraphics*[width=.45\textwidth,angle=270]{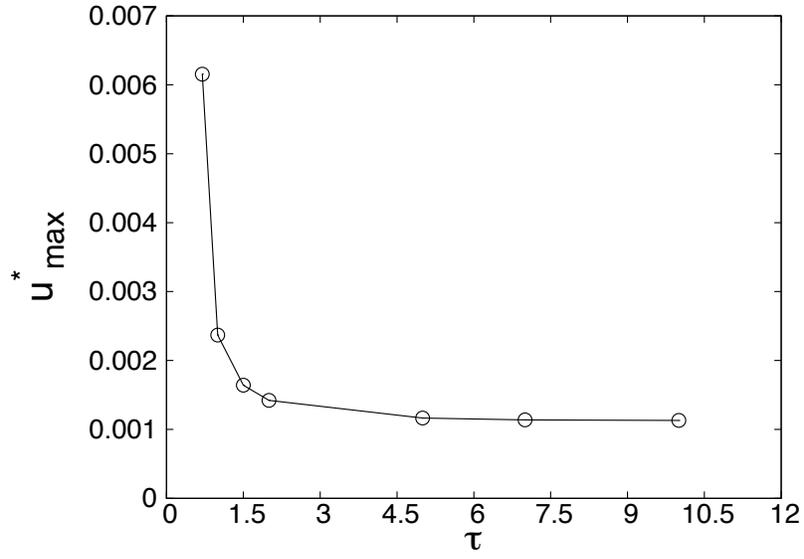}
\caption{The maximum value of the spurious velocities for an equilibrated
droplet is shown as a function of $\tau$
for $\kappa=0.3$ at $T=0.95$ in the OVC case.
Velocity is measured in units of $\Delta x / \Delta t$ 
and time in units of $\Delta t$.
\label{fig:spvel_tau}}
\end{center}
\end{figure}

A very nice feature of the present model is the independence of the
phase diagram on the relaxation time $\tau$. To the best of our
knowledge this was previously observed only in the model proposed in
Ref.~\cite{kupe09}, while
variations in the
equilibrium vapor density up to $50\%$ can be seen in the range $1
\leq \tau \leq 2$ for other models \cite{shan93,he98,luo98,luo00}
(see Fig.~2 of Ref.~\cite{kupe09}).
This feature is illustrated in
Fig.~\ref{fig:dens_tau}, where the equilibrium liquid and vapor
densities at $T=0.95$ are shown, and
holds also at different temperatures.
Here we note that the optimal
values $(N,Q)_{OVC}$ of Table \ref{table} were not found depending
on $\tau$. Finally, the behavior of $u^*_{max}$ as a function of
$\tau$ for the same temperature is presented in
Fig.~\ref{fig:spvel_tau} where a decrease with the relaxation time
is observed.

\section{Zebra phase and shear}

\begin{figure}[!h]
\begin{center}
\includegraphics*[width=.3\textwidth]{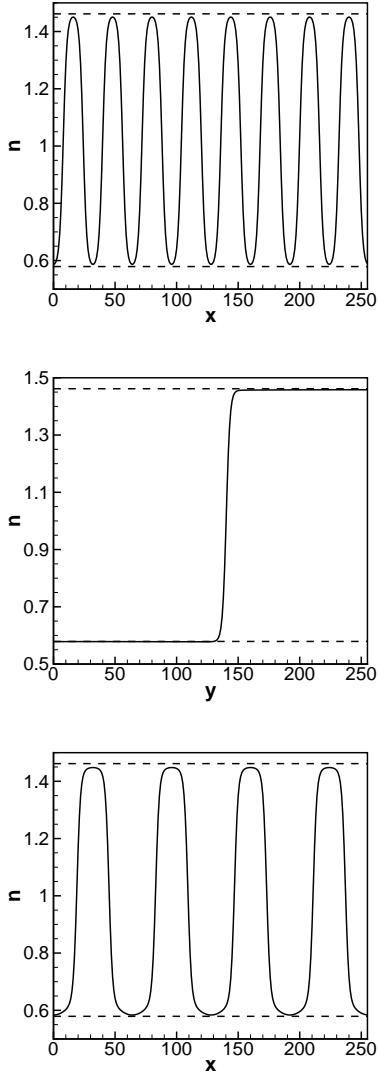}
\caption{(Top) The density profile across the interface of the
zebra phase for a periodic force with $A_0=10^{-4}$ and $\lambda=32$
in the case with $\tau=1$, $\kappa=0.3$ at $T=0.95$.
The measured values for the density in the bulk
are $n_L=1.459906$ and  $n_V=0.580089$.
Dashed lines
represent the
theoretical values  $n_L^{th}$ and  $n_V^{th}$ in a quiescent 
system without periodic
potential at the same temperature (see the caption of Fig.~\ref{fig:interf}).
(Middle) The density profile along the $y$-direction at $x=L/2$
of a system under a periodic force with $A_0=10^{-4}$ and $\lambda=32$
and shear rate $\dot \gamma=10^{-3}$ 
(right bottom panel of Fig.~\ref{fig:zebra}).
(Bottom) The density profile along the $x$-direction at $y=L/2$
of a system under a periodic force with $A_0=10^{-3}$ and $\lambda=64$
and shear rate $\dot \gamma=10^{-3}$ 
(right bottom panel of Fig.~\ref{fig:ampl}).
Time is measured in units of $\Delta t$, lengths in units of
$\Delta x$, and shear rate in units of $\Delta t^{-1}$.
\label{fig:zebrasingle}}
\end{center}
\end{figure}

In presence of the periodic force (\ref{periodicforce}),  a stable
oscillating phase is found with the same period as in
(\ref{periodicforce}) and minimum and maximum values of density
close to the vapor and liquid values at coexistence without
external field, at the temperature under consideration, as shown in
Fig.~\ref{fig:zebrasingle}.
This phase has been called zebra
phase and a detailed study of the phase diagram has been done in
Ref.~\cite{Vink11}. An external flow is expected to distort the zebra
order. We will consider a shear flow imposed by
the top and bottom walls moving with opposite velocities $\pm
\dot{\gamma} L_y/2$ along the $x$-direction. We will show in the following
how the
morphology of the zebra phase is affected by this flow. In a
homogeneous system the shear would give the Couette
profile ${\bf u^*} = (\dot{\gamma} (y-L_y/2), 0)$, as it has been
checked. 
The shear rate will be expressed in units
of $\Delta t^{-1}$.

\begin{figure}[H]
\begin{center}
\includegraphics[width=.8\textwidth]{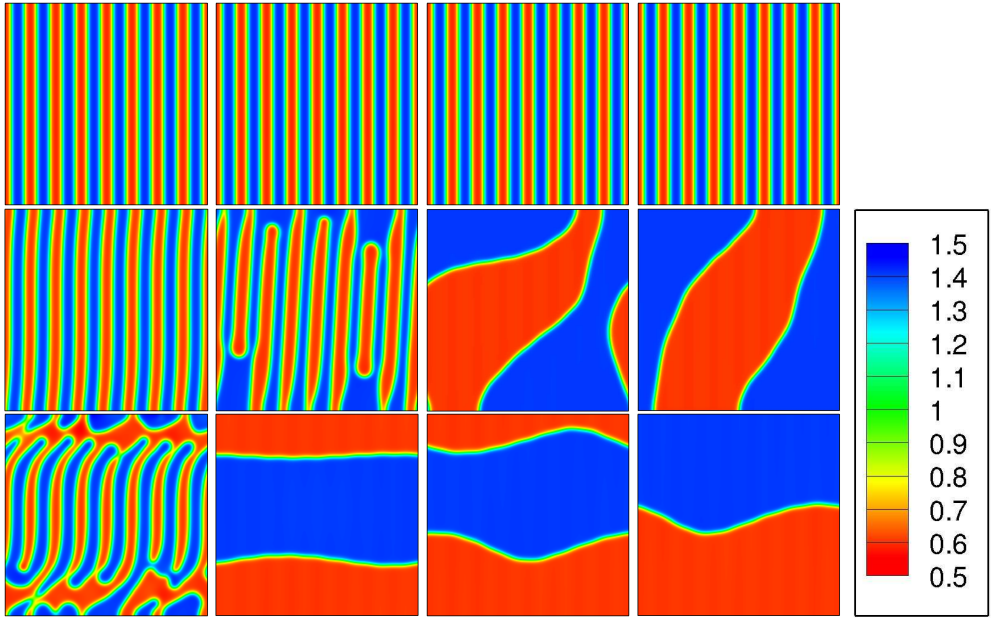}
\caption{Density configurations  at times 
$t = 10^3, 10^4, 10^5, 2
\times 10^5$ (from left to right) for a system of size $256 \times
256$ under a periodic force with $A_0=10^{-4}$ and $\lambda=32$ with
shear rates $\dot{\gamma}=10^{-7}$ (upper row),
$\dot{\gamma} =10^{-4}$ (middle row), and  
$\dot{\gamma} =10^{-3}$
(lower row).
The bulk values of density, when the zebra phase is stable,
correspond to those without shear shown
in the top panel of Fig.~\ref{fig:zebrasingle}.
When the zebra phase is destroyed, liquid and vapor density values
correspond to the ones 
reported in Fig.~\ref{fig:interf}.
Density values are depicted in the color bar at the right of the figure.
Time is measured in units of $\Delta t$, lengths in units of
$\Delta x$, and shear rate in units of $\Delta t^{-1}$.
\label{fig:zebra}}
\end{center}
\end{figure}

Figure \ref{fig:zebra} summarizes our results for the evolution of
the zebra phase at different $\dot{\gamma}$ on a system of size 
$L_x \times L_y = 256 \times 256$ with $T=0.95$, $\tau=1$, and $\kappa=0.3$. 
The system is
prepared through a quenching from a homogeneous state with
symmetric composition at rest and shear is applied from the
beginning of the simulations. The case with $\lambda=32$ and
$A_0=10^{-4}$ is shown. At small values of shear the pattern is
basically unaffected by the flow. At larger values of $\dot{\gamma}$
(middle row in Fig.~\ref{fig:zebra}) the zebra interfaces lose
stability and at long times lamellae are destroyed. Liquid and vapor
phases will coexist separated by an oblique interface that
remains stable at  times much larger than the last one in
Fig.~\ref{fig:zebra}. Small ripples on this interface have the same
periodicity of the external potential. When shear is further
increased (see the case with $\dot{\gamma}=10^{-3}$) lamellae are
destroyed sooner and a stationary state is observed with a single
interface, in average aligned with flow, but with a superimposed
oscillation (see the last snapshot on the right of the lower row
of Fig.~\ref{fig:zebra}).

\begin{figure}[H]
\begin{center}
\includegraphics[width=0.4\textwidth]{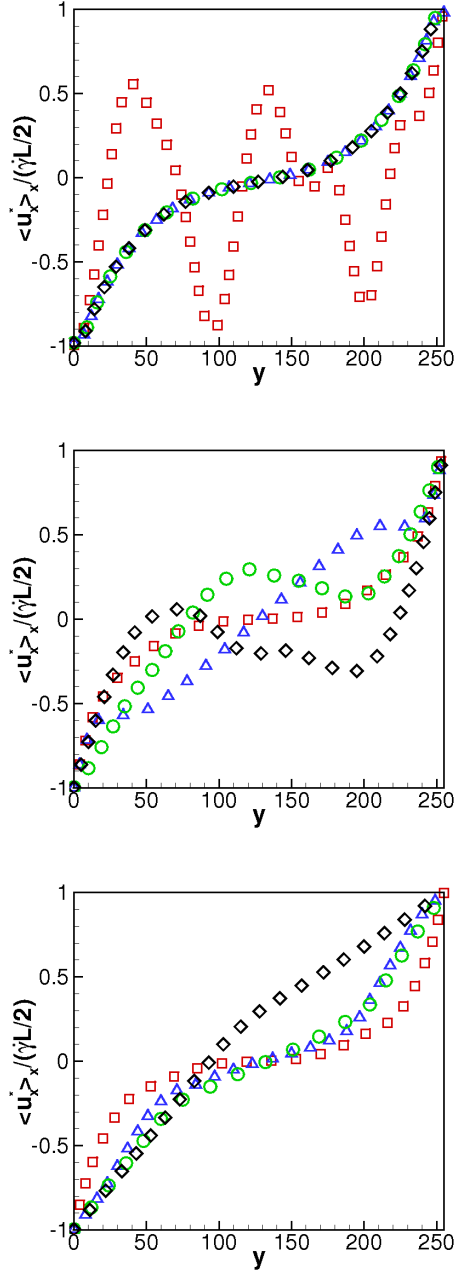}
\caption{Profiles across the system of
the $x$-component of the fluid
velocity averaged along
the $x$-direction
of the patterns shown in Fig.~\ref{fig:zebra}
for $\dot \gamma=10^{-7}, 10^{-4}, 10^{-3}$ (from top to bottom) 
at $t = 10^3$ (red squares), $10^4$ (blue triangles), $10^5$ (green
circles), $2
\times 10^5$ (black diamonds).
The values of $|<u^*_y>_x|/(\dot \gamma L/2)$ are $\lesssim 0.05$
in all the cases.
The maximum values of the spurious velocities are such that
$u^*_{max}/(\dot \gamma L/2) \sim 10^{-4}, 10^{-4}, 10^{-2}$ from top to bottom.
Time is measured in units of $\Delta t$, lengths in units of
$\Delta x$, and shear rate in units of $\Delta t^{-1}$.
\label{fig:zebravel}}
\end{center}
\end{figure}

In all the cases of Fig.~\ref{fig:zebra} shear becomes effective at
times $t$ 
of order $10^4$. This time corresponds to the quantity
$t_S=nL_y^2/(\eta \pi^2)$ which is the leading contribution to the
relaxation time for a shear flow in a simple fluid \cite{schl79}.
The horizontally averaged component $<u^*_x>_x$ of
the fluid velocity is shown in Fig.~\ref{fig:zebravel} at times
corresponding to those  of the snapshots in Fig.~\ref{fig:zebra}.
Figure \ref{fig:zebravel} shows that at the lower values of the
shear rate ($\dot{\gamma}=10^{-7}, 10^{-4}$) a regular linear
velocity profile is not formed (except close to the walls),  even at
the longest simulated time $t=10^6 >> t_S$. 
On the contrary, at the
highest value $\dot{\gamma}=10^{-3}$, the velocity profile along the
flow direction  shows a broken line. The slopes agree with those
expected in a system with two phases of different density
separated by a horizontal interface and given by
$<u^*_x>_x/(\dot{\gamma}L_y/2)=4n_uy/[L_y(n_u+n_l)]-1$ for $0 \leq y
\leq L_y/2$ and $<u^*_x>_x/(\dot{\gamma}L_y/2)=1-4n_l(L_y-y)/[L_y(n_u+n_l)]$
for $L_y/2 \leq y \leq L_y$ where $n_l$ and $n_u$ are the values of
densities in the lower and upper half of the system, respectively
\cite{papa99}.

\begin{figure}[H]
\begin{center}
\includegraphics[width=0.8\textwidth]{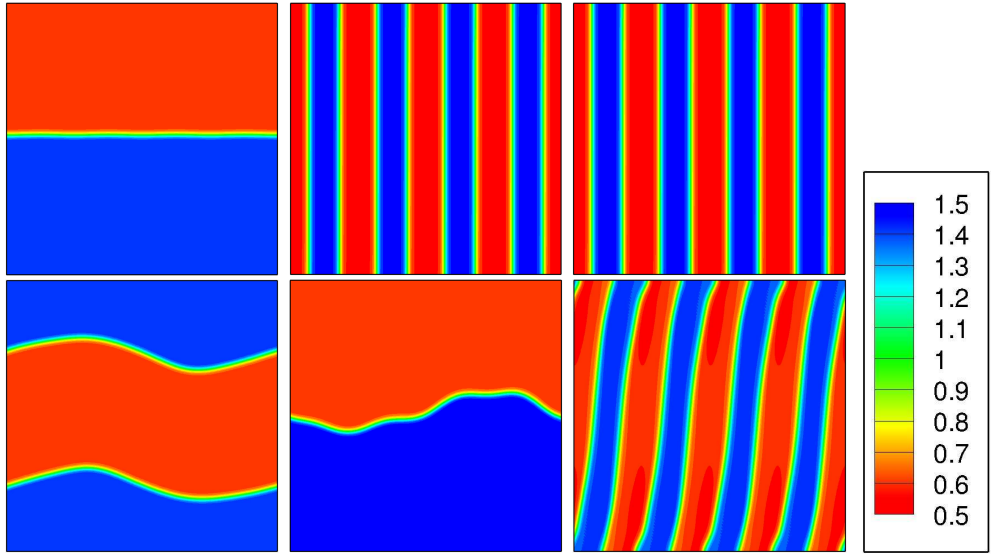}
\caption{Late time density configurations for a system of size $256 \times
256$ under a periodic force with $A_0=10^{-5}, 10^{-4}, 10^{-3}$
(from left to right) and $\lambda=64$ with shear rates
$\dot{\gamma} =10^{-5}$ (upper row) and 
$\dot{\gamma} =10^{-3}$ (lower row).
For the bulk values of the density see the comment in 
Fig.~\ref{fig:zebra}.
Density values are depicted in the color bar at the right of the figure.
Time is measured in units of $\Delta t$, lengths in units of
$\Delta x$, and shear rate in units of $\Delta t^{-1}$.
\label{fig:ampl}}
\end{center}
\end{figure}

By varying $\lambda$ and $A_0$, the pattern evolution, reported
in Fig.~\ref{fig:ampl}, remains similar
to that shown in Fig.~\ref{fig:zebra}. Lamellae with larger
$\lambda$ are more stable against shear.  For example, at 
$\dot{\gamma}= 10^{-4}$ and $A_0 =10^{-4}$ (not shown in 
Fig.~\ref{fig:ampl}), lamellae with
$\lambda=64$ result only slightly distorted close to the walls and
keep their periodicity, differently than in the case of
Fig.~\ref{fig:zebra} with $\lambda=32$ when they are destroyed by
shear. Figure \ref{fig:ampl} also illustrates  the role of different
values of  $A_0$. One observes, for example at
$\dot{\gamma}=10^{-3}$, that the zebra phase remains stable at
$A_0=10^{-3}$ while the fluid completely phase separates at
$A_0=10^{-4}$.

\begin{figure}[H]
\begin{center}
\includegraphics[width=0.99\textwidth]{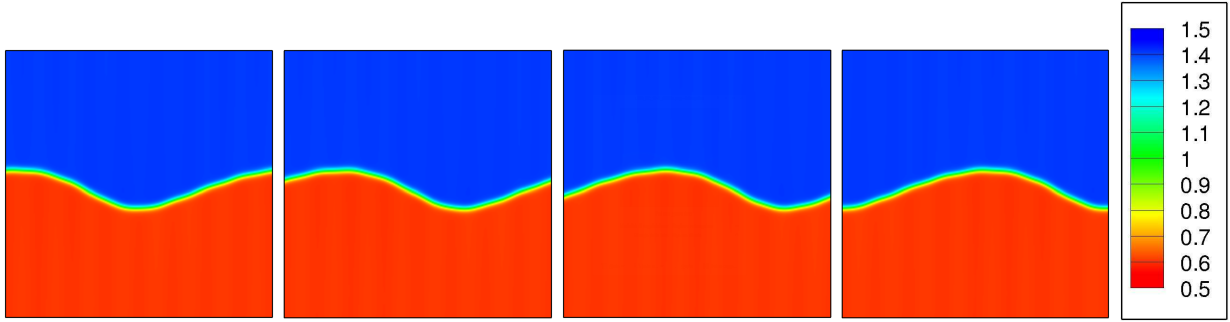}
\caption{Density configurations  at consecutive times 
$t=(4.90, 4.92, 4.94, 4.96) \times 10^5$ (from left to
right) for a system of size $256 \times 256$ under a periodic
force with $A_0=10^{-4}$, $\lambda=32$,  shear rate
$\dot{\gamma} =10^{-3}$. 
The bulk values of the density
correspond to those 
reported in Fig.~\ref{fig:interf}.
Density values are depicted in the color bar at the right of the figure.
Time is measured in units of $\Delta t$, lengths in units of
$\Delta x$, and shear rate in units of $\Delta t^{-1}$.
\label{fig:wave}}
\end{center}
\end{figure}

\begin{figure}[H]
\begin{center}
\includegraphics[scale=0.4]{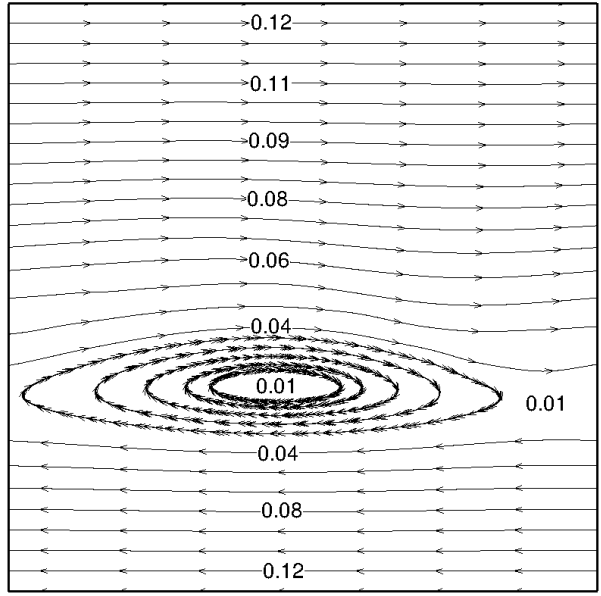}
\caption{The velocity pattern with stream lines and velocity moduli at
selected points
corresponding to the third panel
shown in Fig.~\ref{fig:wave}. For this set of parameters the maximum value
of spurious velocities is 0.002 (see Fig.~\ref{fig:spvel_T}).
Velocities are measured in units of $\Delta x/\Delta t$.
\label{fig:vel}}
\end{center}
\end{figure}

In all the simulations at $T=0.95$, $\tau=1$, $A_0 \leq 10^{-4}$, 
by increasing the shear rate, 
the density field
was observed to  behave as in the bottom right snapshot  of
Fig.~\ref{fig:zebra}. The nature of this state  is better described
in Fig.~\ref{fig:wave}. It is a  peculiar non-equilibrium state,
characterized by a travelling wave on the liquid-vapor interface. 
It may be argued that at large
shear, the oscillations induced by the periodic potential on
the interface are advected by the flow resulting in the interface
dynamics shown in Fig.~\ref{fig:wave}. It is also interesting to see
the behavior of the corresponding  velocity pattern. This is plotted,
for the third panel of Fig.~\ref{fig:wave},
 in Fig.~\ref{fig:vel} which  clearly shows the existence of a
recirculation flow field below the moving
 interface. The magnitude of the fluid velocity in such a pattern
is about one order of 
magnitude larger than spurious currents measured for quiescent droplets
thus confirming the physical origin of the observed structure.
Such a vortex follows the interface wave and 
has been  observed in all the cases where
travelling waves appear as, for example, those with $A_0 =
 10^{-4},10^{-5}$ and $\dot{\gamma}= 10^{-3}$
in Fig.~\ref{fig:ampl}. 

\begin{figure}[H]
\begin{center}
\includegraphics[width=0.99\textwidth,height=0.7\textwidth]{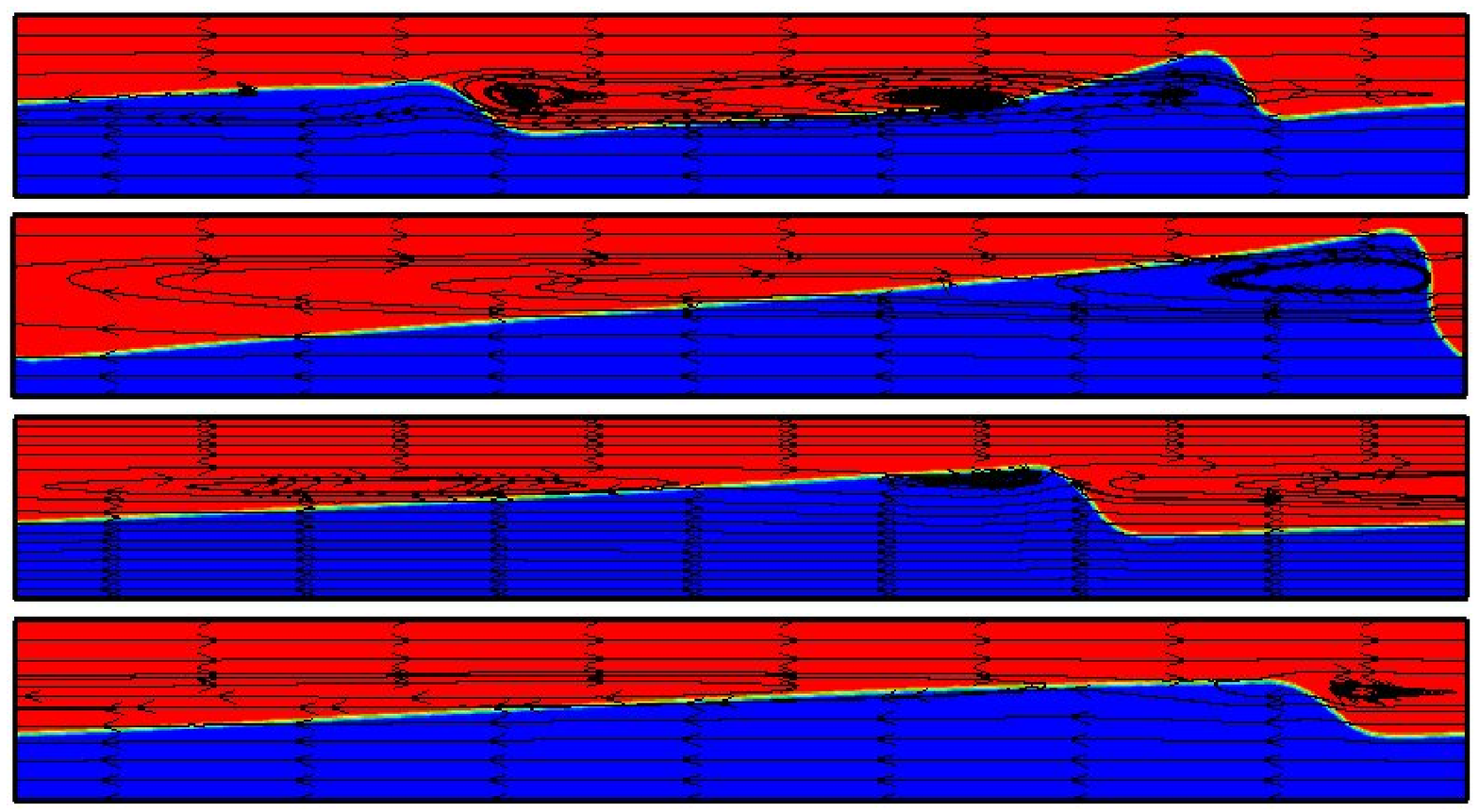}
\caption{Density configurations  with superimposed
velocity patterns at times $t=(15, 19, 23, 100) \times 10^4$ 
(from top to bottom) for a systems of size $2048 \times 256$ 
under a periodic
force with $A_0=10^{-4}$, $\lambda=32$,  and shear rate
$\dot{\gamma} =10^{-3}$.
Density values are depicted in the color bar of Fig.~\ref{fig:wave}.
Time is measured in units of $\Delta t$, length in units of
$\Delta x$, and shear rate in units of $\Delta t^{-1}$.
Aspect ratio of frames is not preserved for a better view.
\label{fig:bigsize}}
\end{center}
\end{figure}

We deeply investigated if and how the wavelength and the amplitude
of the observed wave would depend on the system parameters.
As a first check, the length $L_x$ of the system was systematically increased
assuming the values $512, 1024, 2048$ keeping the width $L_y=256$ fixed
for the case shown in the lower row of Fig.~\ref{fig:zebra}.
A single wave travelling over 
the whole system was always observed at long times.
The evolution of the density patterns when $L_x=2048$, 
leading to the final configuration, is
reported in Fig.~\ref{fig:bigsize} where the typical vortex structure of the
flow field can be also appreciated.
It is interesting to observe in this very long system the presence,
in an intermediate time regime, of two different waves which later
join into a single one indefinitely running over the system. 
Other simulations with
different values of $\lambda, \tau, \kappa, T, A_0, \dot \gamma$ (see below)
confirmed the above pattern with a single wave,
provided the shear rate is strong enough. 

The amplitude $A$ of the interface wave,
computed as the distance between the maximum and the minimum of the wave, 
was found to depend on the
various parameters in a non trivial way. The stationary
values of $A$ are reported
in Table \ref{ampiezza} for several cases.
It results that the dependence of $A$ on the potential wavelength
$\lambda$ and on the surface-tension parameter $\kappa$ is
negligible. 
$A$ grows with the length $L_x$ of the system, saturating at the largest
value of $L_x$, and with the
shear rate $\dot \gamma$. 
On the contrary, we observe that $A$ decreases
when increasing the relaxation time $\tau$, the potential amplitude $A_0$, and
the temperature $T$. Actually, it was found that the interface wave survives
down to $T=0.91$, where its amplitude is about a half of the system width, 
but dissolves for $T \leq 0.90$ where a flat horizontal 
interface separates liquid from vapor.

We also wanted to analyze the occurrence of the non-equilibrium
state described above by using different initial conditions. We
chose the most obvious initial state with a flat liquid-vapor
interface parallel to the flow direction and studied the evolution
of the system for different cases previously considered. 
We found the same behavior with a travelling wave
as in Figs.~\ref{fig:wave}-\ref{fig:vel}
at large shear rates
and we concluded  that this picture
is a general effect of the simultaneous presence of shear and
periodic potential, independently of the choice of initial
conditions. 
We also checked that the observed waves 
do not correspond to transient regimes. Indeed,
the time evolution
of the total interface length 
for systems of different sizes and initial conditions always shows
a plateau at long times.
On the other side, for  
$\dot{\gamma} \le
10^{-4}$ (at $A_0=10^{-4}$ and $\lambda=32$), at large times, 
flat interfaces
were found. 
We did not find evolutions resulting in oblique
interfaces like those at $\dot{\gamma}=10^{-4}$ in
Fig.~\ref{fig:zebra}.
Therefore,  differently than the case
at larger shear rate, at small shear the
stationary patterns
depend on the initial state of the system.

Finally, we checked whether it would be possible to explain the
presence of the  interface waves  as due to a sort of
Kelvin-Helmholtz instability \cite{chan61}. This is an interface
instability  that may occur   when  two fluids with different density
have different
tangential velocities.
By varying  the shear rates or other parameters, we never observed
the wave amplitude to grow making the interface unstable
as in the Kelvin-Helmholtz instability that,
we concluded, cannot be referred  to our case.

\section{Conclusions}
\label{sec:conclusion}

In this paper we have shown the effects of a shear flow on the
density-oscillating ``zebra'' phase appearing in liquid-vapor systems
in presence of a periodic potential. For this purpose, we have
introduced  a LBE with Gauss-Hermite quadrature
in order to simulate van der Waals fluids. 
The advantages of this approach are that i) it is based 
on a systematic discretization of LBE, ii) spurious terms  
are avoided in the continuum equations 
(except for the $u^{*3}$ term in the Navier-Stokes equation), and iii) the
equilibrium phase diagram does not depend on the relaxation time of the LBE.
Higher isotropy derivatives have been considered in order to reduce spurious 
velocities. However, our results show that a larger reduction 
has been obtained by other
approaches \cite{wagn07,nour02}.

We considered a
region of the phase diagram where the ``zebra'' phase is stable.
 Shear was applied with flow perpendicular to the lamellae.
Striped patterns with larger period or in a deeper  potential, in general,
were   found less distorted by the flow. At high shear rates the
``zebra''  phase  becomes unstable and the system completely separates
into liquid and vapor regions. This state, however, is far from
being trivial. It is  characterized by travelling interface
waves with steady total interface length at late times and by a
recirculation pattern of the velocity field. 
Also in very large systems we  always observed, at late times, 
a single perturbation moving over the lattice
and one could ask about the origin of this  behavior, 
in absence of any analytical proof
of this instability.
By increasing the length $L_x$, keeping fixed the other parameters,
the energy injected into the system, due to the applied shear, increases 
with $L_x$.
The increased amount of  energy has to be organized 
in the velocity field and in the interface pattern
and, in principle, more waves could also be   expected.
However, our results show 
that the injected energy always prefer to be organized in 
a single wave with the  wavelength of the perturbation defined solely by 
the length $L_x$ of the
channel. The corresponding velocity pattern is characterized by a single 
vortex -
even  if more waves and vortices are present at 
intermediate times.  
This behavior can be attributed to the fact 
that: 
1. the external driving induces extended  perturbations 
spanning over 
the whole system; 2. 
different waves, interacting each other,
have  an interfacial energy cost so that 
 the coalescence  into  a single wave, when periodic boundary conditions 
(PBC) are applied,
as shown in the sequence of configurations of Fig. 13, is reasonable. 
Since the use of PBC appears to be determinant in the above results,
we also observe that the simulated system with 
periodic boundary conditions corresponds to consider a plane, normal to the 
symmetry axis, of the experimental setup (Couette cell) made of two co-axial 
counter-rotating cylinders.

Finally, we ask whether we can expect quantitative agreement between
simulations and physical systems. We take the surface tension $\sigma$, the 
viscosity of the liquid phase
$\eta_L$ and the interface width $\xi$ of the lattice Boltzmann (LB) model 
to correspond to the physical values via the relations
$\xi^{LB}\Delta x^{ph}=\xi^{ph}$, $\sigma^{LB}\Delta m^{ph}/(\Delta t^{ph})^2=
\sigma^{ph}$, and $\eta^{LB}\Delta m^{ph}/(\Delta t^{ph} \Delta x^{ph})=
\eta^{ph}$. This defines physical length, time, and
mass scales.
In the case
of mixtures of spherical silica colloids and PDMS polymers in solvent
\cite{hoog99} $\sigma^{ph}=3 \times 10^{-6} N/m$,
$\eta_L^{ph}=0.097 Pa \; s$, and $\xi^{ph} \simeq 10^{-8}m$. 
The interfacial thickness is about the size of the
hydrodynamic radius of colloids and of the radius of gyration of
polymers.
For the values of $\xi^{LB}, \sigma^{LB}, \eta^{LB}$
used in the present paper, the physical scales are 
$\Delta x^{ph}=1.7 \times 10^{-9}m$ and $\Delta t^{ph}=1.2 \times 10^{-6} s$.
With these values we can predict travelling waves to be
observable at shear rates 
$\dot \gamma^{ph}=10^{-3}/\Delta t^{ph}=8.3 \times 10^2 s^{-1}$
with $\lambda^{ph}=64 \Delta x^{ph}=1.1 \times 10^{-7} m$ \cite{nota}.
We also expect that
systems in $d=3$ will exhibit a still richer morphology but we leave this
case for a future study.

\newpage

\begin{table}[h]
\begin{tabular}{|l|c|c|c|c|}
\hline $\;\;T$ & $(N,Q)_{OVC}$&$(N,Q)_{ODC}$\\
\hline 0.83&(0.3,2.1)&(0.3,2.1)\\
\hline 0.85&(0.3,2.3)&(0.3,2.3)\\
\hline 0.87&(0.3,2.5)&(0.3,2.5)\\
\hline 0.89&(0.3,2.6)&(0.3,2.7)\\
\hline 0.91&(0.3,2.5)&(0.4,1.7)\\
\hline 0.93&(0.3,1.9)& (0.4,1.9)\\
\hline 0.95&(0.3,2.0)&(0.4,2.1)\\
\hline 0.97&(0.3,0.0)&(0.0,0.0)\\
\hline 0.98&(0.3,3.9)&(0.3,3.9)\\
\hline 0.99&(0.5,2.1)&(1.0,1.0)\\
\hline
\end{tabular}\\[10 pt]
\caption{Optimal values of $N$ and $Q$ in
(\ref{xstencil})-(\ref{laplstencil}) for the OVC and ODC cases
with $\kappa=0.3$.}
\label{table}
\end{table}

\newpage

\begin{table}[H]
\scalebox{0.85}{
\begin{tabular}{|l|c|c|c|c|c|c|c|}
\hline $\;\;T$&$\lambda$&$\tau$&$\dot{\gamma}$&$A_0$&$\kappa$&$L_x\times L_y$&$A/L_y$\\
\hline 0.95&32&1&$10^{-3}$&$10^{-4}$&0.3&$256\times 256$&0.172\\
\hline 0.95&64&1&$10^{-3}$&$10^{-4}$&0.3&$256\times 256$&0.172\\
\hline 0.95&32&1&$10^{-3}$&$10^{-4}$&0.3&$512\times 256$&0.316\\
\hline 0.95&64&1&$10^{-3}$&$10^{-4}$&0.3&$512\times 256$&0.316\\
\hline 0.95&32&1&$10^{-3}$&$10^{-4}$&0.3&$1024\times 256$&0.409\\
\hline 0.95&64&1&$10^{-3}$&$10^{-4}$&0.3&$1024\times 256$&0.409\\
\hline 0.95&32&1&$10^{-3}$&$10^{-4}$&0.3&$2048\times 256$&0.417\\
\hline 0.95&64&1&$10^{-3}$&$10^{-4}$&0.3&$2048\times 256$&0.417\\
\hline 0.95&32&1&$10^{-3}$&$10^{-5}$&0.3&$256\times 256$&0.178\\
\hline 0.95&64&1&$10^{-3}$&$10^{-5}$&0.3&$256\times 256$&0.178\\
\hline 0.95&32&1&$10^{-3}$&$10^{-5}$&0.3&$512\times 256$&0.335\\
\hline 0.95&64&1&$10^{-3}$&$10^{-5}$&0.3&$512\times 256$&0.335\\
\hline 0.95&32&1&$10^{-3}$&$10^{-5}$&0.3&$1024\times 256$&0.504\\
\hline 0.95&64&1&$10^{-3}$&$10^{-5}$&0.3&$1024\times 256$&0.504\\
\hline 0.95&32&1&$10^{-3}$&$10^{-4}$&0.2&$256\times 256$&0.172\\
\hline 0.95&32&1&$10^{-3}$&$10^{-4}$&0.5&$256\times 256$&0.172\\
\hline 0.95&32&2&$10^{-3}$&$10^{-4}$&0.3&$256\times 256$&0.112\\
\hline 0.95&64&2&$10^{-3}$&$10^{-4}$&0.3&$256\times 256$&0.112\\
\hline 0.95&32&2&$10^{-3}$&$10^{-4}$&0.3&$512\times 256$&0.219\\
\hline 0.95&64&2&$10^{-3}$&$10^{-4}$&0.3&$512\times 256$&0.219\\
\hline 0.95&32&3&$10^{-3}$&$10^{-4}$&0.3&$256\times 256$&0.00782\\
\hline 0.95&64&3&$10^{-3}$&$10^{-4}$&0.3&$256\times 256$&0.0234\\
\hline 0.95&32&3&$10^{-3}$&$10^{-4}$&0.3&$512\times 256$&0.166\\
\hline 0.95&64&3&$10^{-3}$&$10^{-4}$&0.3&$512\times 256$&0.166\\
\hline 0.91&32&1&$10^{-3}$&$10^{-4}$&0.3&$256\times 256$&0.255\\
\hline 0.91&64&1&$10^{-3}$&$10^{-4}$&0.3&$256\times 256$&0.255\\
\hline 0.91&32&1&$10^{-3}$&$10^{-4}$&0.3&$512\times 256$&0.397\\
\hline 0.91&64&1&$10^{-3}$&$10^{-4}$&0.3&$512\times 256$&0.397\\
\hline 0.91&32&1&$10^{-3}$&$10^{-4}$&0.3&$1024\times 256$&0.496\\
\hline 0.91&64&1&$10^{-3}$&$10^{-4}$&0.3&$1024\times 256$&0.496\\
\hline
\end{tabular}
}
\vskip 2.cm
\caption{The amplitude $A$ of interface waves is reported for
different system parameters.}
\label{ampiezza}
\end{table}

\newpage

\newpage

\newpage

\newpage

\newpage

\newpage

\newpage

\newpage

\newpage

\newpage

\newpage

\newpage

\newpage

\begin{thebibliography}{99}
\bibitem{nota}
Observe that, as well known \cite{yeo06}, 
it is not possible to choose all the variables to take
the physical values still being able to perform stable and affordable runs
since simulations of realistic experimental
setups would require very large lattices
(about $10^5$ nodes per direction!).
\end{thebibliography}
\end{document}